\documentclass[%
 reprint,
 amsmath,amssymb,
 aps,
]{revtex4-1}

\usepackage{graphicx}
\usepackage{dcolumn}
\usepackage{bm}

\usepackage{color}
\usepackage{graphics}
\usepackage{siunitx} 
\usepackage{float}

\usepackage{comment}

\newcommand{\A}[0]{\mathrm{A}}
\newcommand{\B}[0]{\mathrm{B}}
\newcommand{\E}[0]{\mathrm{E}}

\begin{document}


\title{{Finite size mediated radiative coupling of lasing plasmonic bound state in continuum}}


\author{Benjamin O. Asamoah$^1$}
\author{Marek Nečada$^{1,2}$}
\author{Wenzhe Liu$^3$}
\author{Janne Heikkinen$^1$}
\author{Sughra Mohamed$^1$}
\author{Atri Halder$^1$}
\author{Heikki Rekola$^1$}
\author{Matias Koivurova$^{4,5}$}
\author{Aaro I. V\"{a}kev\"{a}inen$^2$}
\author{P\"{a}ivi T\"{o}rm\"{a}$^2$}
\author{Jari Turunen$^1$}
\author{Tero Set\"{a}l\"{a}$^1$}
\author{Ari T. Friberg$^1$}

\author{Lei Shi$^6$}\email{lshi@fudan.edu.cn}
\author{Tommi K. Hakala$^1$}\email{tommi.hakala@uef.fi}

\affiliation{$^1$Institute of Photonics, University of Eastern Finland, P.O. Box 111, FI-80101 Joensuu, Finland}
\affiliation{$^2$Department of Applied Physics, Aalto University, FI-00076 Aalto, Finland}
\affiliation{$^3$Department of Physics, The Hong Kong University of Science and Technology, Clear Water Bay, Kowloon, Hong Kong, China}
\affiliation{$^4$Tampere University, Tampere Institute for Advanced Study, 33100 Tampere, Finland}
\affiliation{$^5$Tampere University, Faculty of Engineering and Natural Sciences, 33720 Tampere, Finland}
\affiliation{$^6$State Key Laboratory of Surface Physics, Key Laboratory of Micro- and Nano-Photonic Structures (Ministry of Education) and Department of Physics, Fudan University, Shanghai, China}



\begin{abstract}
{Radiative properties of lasing plasmonic bound state in continuum are analyzed. The topological charge of the lasing signal is analyzed in the far field as well as in the source plane of the finite sized plasmonic lattice. The physical mechanism enabling the coupling of the BIC to radiation continuum is identified. We show that while the BICs have their origin in multipolar resonances, their far-field radiation properties are governed by the position dependent dipole moment distribution induced by the symmetry breaking in a finite plasmonic lattice. Remarkably, this dipole-moment enabled coupling to radiation continuum maintains the essential topological features of the infinite lattice BICs.} 
\end{abstract}

\maketitle


\textit{Introduction --} 
Plasmonic nanoparticle lattices are a versatile platform for band structure engineering. Early works have shown remarkably narrow linewidths as compared to single particle resonances, mainly due to reduction of the radiative losses~\cite{zou_silver_2004,garcia_de_abajo_colloquium_2007,kravets_extremely_2008,auguie_collective_2008,Rodriguez2011,wang_rich_2018}. This has enabled, for instance, lasing and Bose-Einstein condensation at the high symmetry points of the lattice \cite{Zhou2013, hakala_bose-einstein_2018}. More recently, multimode lasing as well as multiple output beams with specific polarization properties have been reported \cite{wang_structural_2018, Guo2019, Asamoah}. The feedback mechanism is based on counter propagating in-plane radiation fields in the principal lattice directions, whose relative phase differences can attain only specific values due to periodic boundary conditions (in a square lattice typically either 0 or $\pi$).

Recently, lasing action from so called bound states in continuum (BICs) have been demonstrated in photonic crystals \cite{Kodigala2017, rybin2017optical, Huang1018, mohamed2020topological}. They co-exist in the same energy range with radiative continuum modes, but nevertheless exhibit infinite radiative Q factors. This is due to either symmetry mismatch of their polarization properties with the radiation fields, or simultaneous destructive interference of all the radiation channels \cite{SoljacicBIC, SoljacicRev}. The first type, so called symmetry protected BICs, exhibit a polarization singularity (typically in a high symmetry point of the reciprocal lattice) and winding of the polarization vector around the singularity, which is characterized by a conserved quantity known as a topological charge \cite{PhysRevLett.113.257401}. A few indications of BIC lasing in hybrid plasmonic-waveguide systems have been reported, but no explicit connection to BICs or non-trivial topology was established \cite{Guan2020, Guan2020_2}. Quasi-BIC lasing in plasmonic quadrumer nanoparticle array was considered in~\cite{Heilmann2022}, without a thorough study of the outcoupling mechanism of the radiation.

{As the main result of the present work, we reveal the physical mechanism that allows the coherent coupling of the finite size BIC to the radiation continuum. We show that while the BIC mode is supported by the multipolar plasmonic resonance of the particle, its far-field radiation properties are governed by the position dependent dipole moment distribution over the finite lattice. Remarkably, this dipole-moment enabled coupling to radiation continuum maintains the topological features of the initially subradiant quadrupole excitation. As a side result, we present a first demonstration of a transition from topologically trivial dipolar lasing to multipolar BIC lasing in a plasmonic lattice.}

\textit{Results --}
Our samples consist of cylindrical Au nanoparticles in square lattices with a periodicity $p = \SI{580}{nm}$ and height of 50 nm. Various samples were fabricated with diameters $d$ ranging from 120 to 200 nm in steps of 20 nm. The lattices were overlaid with a gain medium having 25 mM concentration of IR-792 fluorescent molecules in BA:DMSO (2:1) solution. Optical pumping was carried out with a pulsed laser (\SI{150}{fs} pulse duration, \SI{790}{nm} central wavelength, \SI{1}{kHz} repetition rate). Mode and field pattern calculations were done with an implementation of the multiple-scattering T-matrix method, QPMS \cite{necada_multiple-scattering_2021,QPMS}. The single-nanoparticle T-matrices were calculated with the null-field method for cylindrical scatterers of height $h = 50$~nm, using an isotropic background medium with a refractive index $n = 1.52$, and the Drude-Lorentz model for gold \cite{rakic_optical_1998}, for details, see Supplemental Material.

\begin{figure}[ht!]
\centering
\includegraphics[width=0.9\columnwidth]{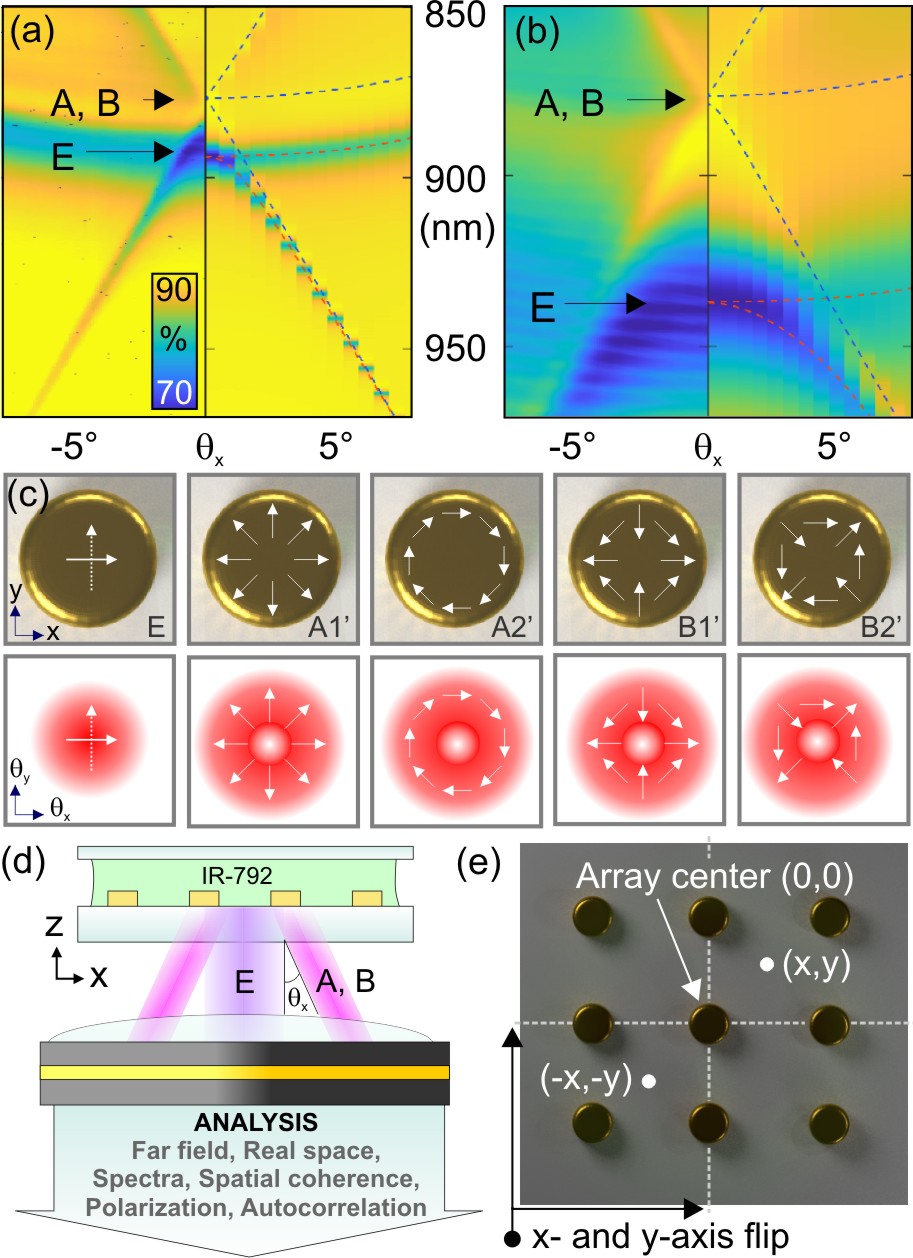}
\caption{(a) Left: Measured angle resolved transmission spectrum for $p = \SI{580}{nm}$, $d = \SI{120}{nm}$ sample. Right: FDTD simulations revealing the spectral position of the subradiant $\A$ and $\B$ modes. (b) Similar to (a), but for $d = \SI{160}{nm}$. Notably the E mode energy depends sensitively on the particle diameter. (c) The  irreducible representations of the modes supported by $D_{4h}$ symmetry. Top: Nanoparticle  polarizations for each mode in an infinite lattice, right: far field polarization properties. (d) Schematic of the sample overlaid with index matched liquid gain medium. The angle, position, wavelength and polarization resolved imaging of the lasing emission is carried out to characterize the observed modes. (e) WFI is used to map polarization resolved spatial coherence and phase correlations between the points $(x,y)$ and $(-x,-y)$ and in particular at the opposite edges of the finite lattice. 
}
\label{fig:intro}
\end{figure}

To recover the band structure, transmission measurements were carried out. The left side of Fig.~\ref{fig:intro}~(a) shows the measured angle resolved unpolarized transmission spectrum for a square lattice with \SI{120}{nm} particle diameter. At \ang{0}, the transmission of the lower energy branch (indicated by $\E$) is 70~\%, while the higher energy branch ($\A$, $\B$) has over 90~\% transmission, indicative of subradiant character of the modes at that energy.  Importantly, several subradiant BIC modes can be found from the vicinity of the higher energy branch. The right side of Fig.~\ref{fig:intro}(a) shows the finite-difference time-domain (FDTD) simulated transmittances together with the calculated mode energies (dashed lines). In Fig.~\ref{fig:intro}(b) the particle diameter is increased to 160 nm. Notably, the radiant lower energy E mode redshifts strongly, while the subradiant $\A$ and $\B$ modes remain at a near constant energy.

The overall square lattice can be classified to the $D_{4h}$ symmetry group \cite{dresselhaus_group_2008}. In the case of an infinite lattice (i.e., periodic boundary conditions), the irreducible representations for each mode obtained from standard group theory \cite{bradley_mathematical_1972} can be employed to recover the polarization pattern within the nanoparticles, see Fig.~\ref{fig:intro}(c) top row. For conciseness, we focus on the modes whose in-plane electric field component is dominant. Fig.~\ref{fig:intro}(c) bottom row shows the corresponding far field patterns.

The symmetry of the E modes (whose polarizations are denoted by continuous and dashed arrows on the left column of Fig.~\ref{fig:intro}(c)) indicates they are degenerate, and thus their superpositions will also be eigenmodes of the system. In contrast, $\A$ and $\B$ modes are nondegenerate and exhibit revolving polarization patterns around the center of the particle. We note that for both $\A$ modes, a closed loop in counterclockwise direction around the center of the nanoparticle is accompanied with a counterclockwise rotation of the polarization direction. In contrast, for $\B$ modes the polarization rotation is opposite (clockwise). The topological charge associated to the winding of the polarization is thus +1 for A modes and -1 for B modes. Notably, for E modes the polarization distribution is antisymmetric (odd) with respect to $C_{2z}$ symmetry operation ($\pi$ rotation around the z axis) while $\A$ and $\B$ modes modes are symmetric (even). Thus, all $\A$ and $\B$ modes are associated with zero dipole moment at the center of the particle, unit cell-averaged zero net dipole moment, and consequently, negligible radiation. 


We analyze both far field and source plane intensity and polarization distributions to identify the modes, Fig.~\ref{fig:intro} (d), and employ a custom built wavefront folding interferometer (WFI) to study the spatial coherence properties of the modes, Fig.~\ref{fig:intro}(e) \cite{Koivurova:19,Halder:20}. In particular, the WFI interferes a source plane radiation field with its replica that is flipped with respect to $x$- and $y$-axis, allowing us to analyze correlations of the radiation at the opposite edges of the lattice.

\begin{figure}
    \centering
    \includegraphics[width=\columnwidth]{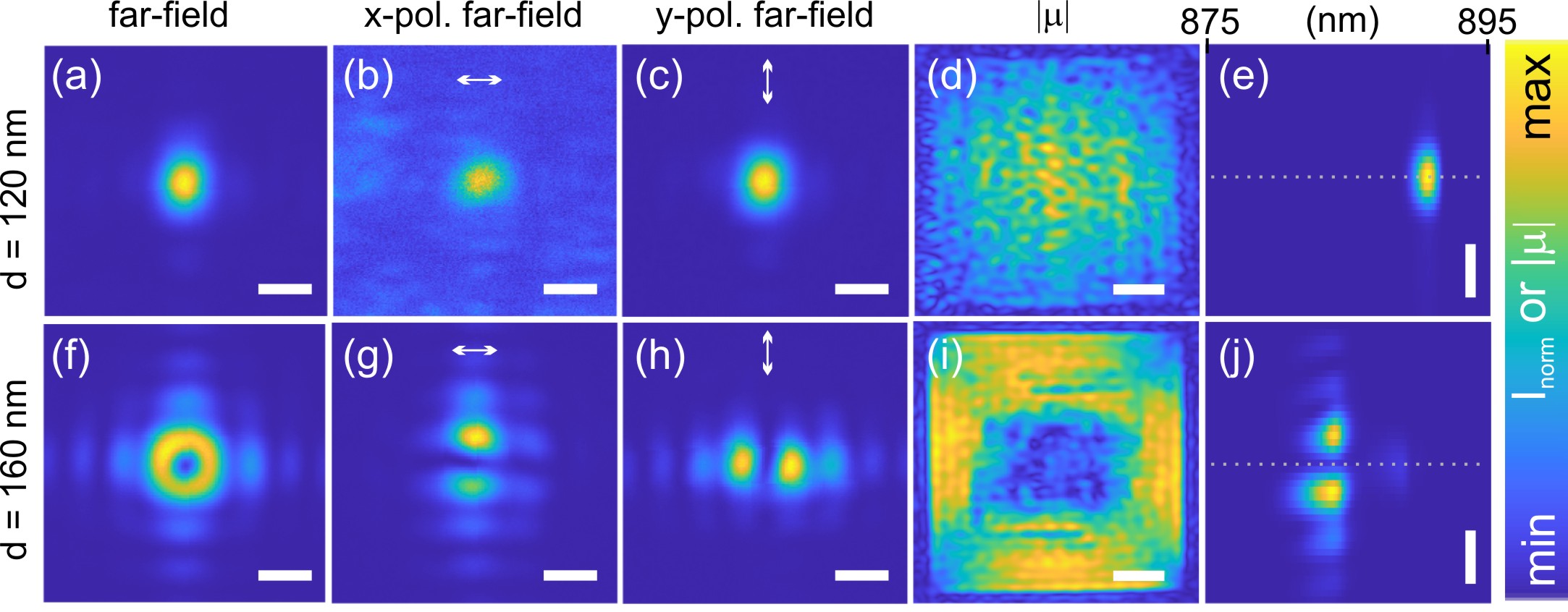}
    \caption{Emission analysis for two samples with $d = \SI{120}{nm}$ (a–e) and $d = \SI{160}{nm}$ (f–j). The unpolarized (a,f), $x$-polarized (b,g) and $y$-polarized (c,h) far field intensity distributions. The spatial degree of coherence, $\lvert \mu \rvert$, at the source plane of the two samples (d,i). The maximum coherence reaches 0.6 and 0.8 in (d) and (i), respectively. (e,j) wavelength and angle resolved emission. Scale bars: \ang{1} in (a–c) and (f–h). In (d,i), scale bar is \SI{20}{\micro m}.  In (e,j), the scale bar is again \ang{1}, but applies only in vertical direction. The dashed lines indicate $\theta_y = 0$.
    }
    \label{fig:exp}
\end{figure}

Figure~\ref{fig:exp}(a) shows the far field emission pattern of the \SI{120}{nm} diameter particle array with a pump fluency of $1.2P_\mathrm{th}$, where $P_\mathrm{th}$ is the threshold fluency. First, we note that the emission is directed normal to the sample with a pronounced single intensity maximum. Figures 2 (b,c) imply that the emission contains both horizontal and vertical polarization components. The spatial coherence properties at the source plane were studied with the WFI, see Fig.~2 (d,i), where the magnitude of the complex valued degree of spatial coherence, $\lvert \mu \rvert$ is presented (see also Supplemental Material, section Correlation functions). The spatial coherence reaches a value $\lvert \mu \rvert=0.6$ in the center of the lattice. Previously, such spatial coherence and beam patterns have been associated with the so-called bright mode of the plasmonic lattice, a hybrid composed of diffracted orders of the lattice and dipolar plasmonic excitations in each particle \cite{Hoang2017}. 

For 160 nm particles, an entirely different intensity pattern is observed, see Fig.~\ref{fig:exp}(f). In particular, a donut shaped beam is observed with negligible emission to the normal direction of the sample. Horizontally polarized light is observed at the top and bottom parts, while vertically polarized light is observed on the left and right sides of the beam, see Figs.~\ref{fig:exp} (g,h). Additionally, a drastic change in the spatial coherence of the sample is observed: the center of the lattice has very low values of $\lvert \mu \rvert$, while the edges have values of 0.8. Importantly, the high $\lvert \mu \rvert$-values observed at the edges of the lattice imply coherent emission in two dimensions, such that even the points $(x,y)$, $(-x,-y)$ (as indicated in Fig.~\ref{fig:intro}(e)) have a phase correlation. 

We note that the evolution of both real space coherence properties and far field intensity patterns are accompanied with changes in the spectral properties. For the 120 nm sample, the angle and wavelength resolved spectra in Fig.~\ref{fig:exp}(e) reveal one intensity maximum at around 890 nm and $\theta_y$ = 0. However, for the 160 nm sample, the lasing signal is blue shifted by about 7 nm and also exhibits a beating pattern with a \textit{minimum} at $\theta_y = 0$ (Fig.~\ref{fig:exp}(j)). We carried out measurements with other diameters (140, 180 and 200 nm) as well, which showed essentially similar behaviour with 160 nm case (see Supplemental Material Fig.~1). {Notably, while the increase of the particle size did not change the polarization properties of the lasing mode, the lasing threshold decreased and wavelength redshifted with increasing particle size, in agreement with our T-Matrix simulations (see Supplemental Material Fig.~5).} Next, we will focus in detail on the properties of the 200 nm sample, whose emission was particularly intense and therefore easy to characterize.

\begin{figure} \centering \includegraphics[width=\columnwidth]{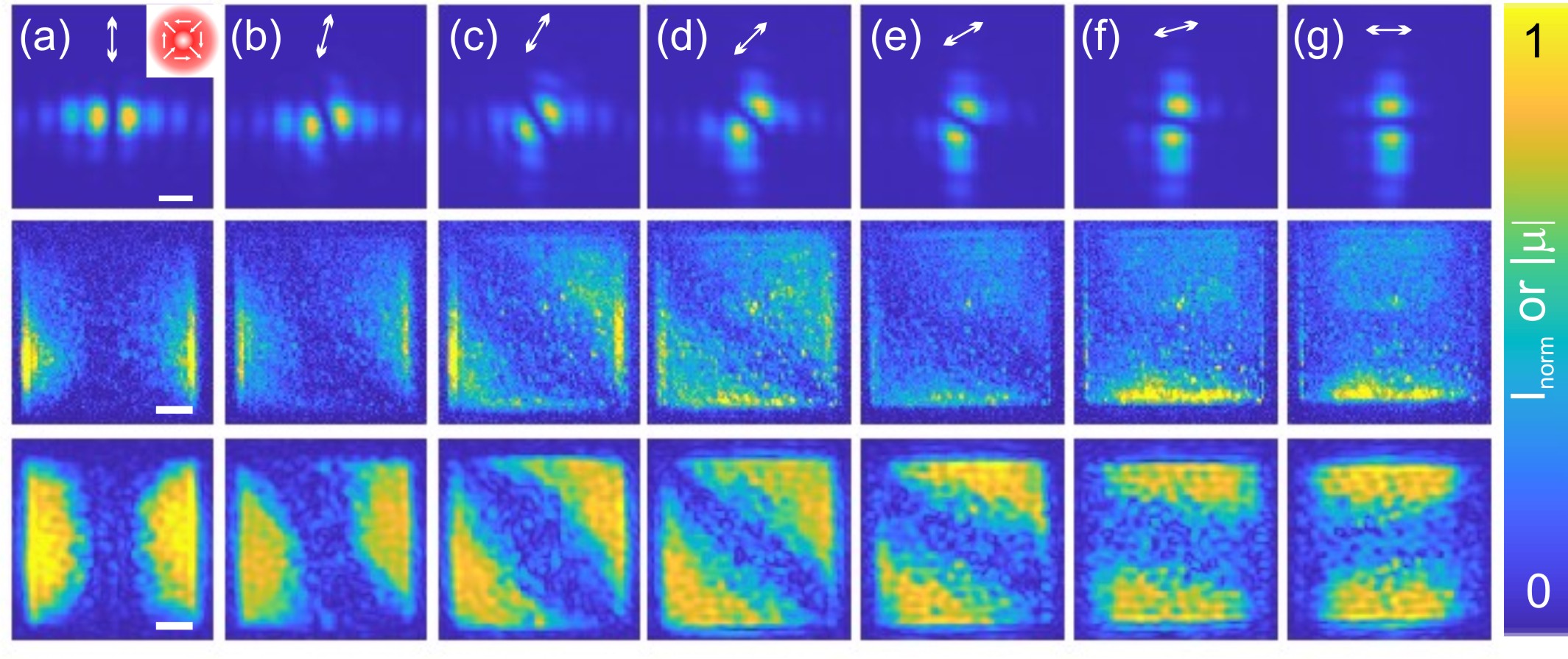}
  \caption{Emission analysis for a sample with $d = \SI{200}{nm}$. First row: polarization resolved far field emission. For each column, the lasing emission was filtered with a linear polarizer whose orientation was rotated in clockwise direction in steps of \ang{15} (indicated by the arrow). The inset in (a) shows the associated polarization. Second row: source plane intensity distributions with the same polarizer orientations as in the first row. Third row: the degree of spatial coherence, $\lvert \mu \rvert$, obtained by WFI measurements with the same polarizer orientations as in the first row. For the first row, the scale bar is \ang{1}, for the second and third rows
  \SI{20}{\micro m}.  } \label{fig:ffangles} \end{figure}

The top row of Fig.~\ref{fig:ffangles} represents the far field images of the emission from the 200 nm sample with consecutive \ang{15} clockwise rotations of a linear polarizer. Similar to the 160 nm sample, the 200 nm sample exhibits horizontally aligned intensity maxima when the polarizer is vertically oriented. As the polarizer is rotated clockwise, the bright lobes rotate counterclockwise. The inset of Fig.~\ref{fig:ffangles}(a) shows the associated polarization pattern. We also carried out time resolved coherence measurements for the far field emission. The coherence approaches unity at zero time delay with a FWHM of 1.5 ps, indicative of ultrafast dynamics of plasmonic BIC lasers (for details of the measurement and experimental data, see Supplemental Material Fig.~2.


Fig.~\ref{fig:ffangles} middle row presents the polarization resolved real space intensity distributions. We note that the overall intensity patterns have their maxima at the edges of the structure and they rotate counterclockwise when the polarizer is rotated clockwise. The measured real space coherence properties are shown on the bottom row. Notably, the sample exhibits spatial coherence values approaching unity at the opposite sides of the lattice for each polarization angle. Further, they rotate in counterclockwise direction along with the intensity maxima. This indicates that there are well defined phase correlations in the dipole moment distribution over the entire lattice. The corresponding interference fringe patterns can be found from Supplemental Material Fig.~3.

Importantly, our results indicate an intimate connection between the far field
polarization properties of the lasing BIC and the location specific dipole
moment distribution at the source plane of the finite size plasmonic lattice:
While the far field polarization distribution exhibits a winding of $2\pi$
around the polarization vortex at the $\Gamma$ point, the real space
polarization distribution exhibits a similar winding around the center of the
entire lattice. Next, we will establish a connection between 1) topology
governed polarization winding in each unit cell of an infinite lattice, 2) real
space dipole distribution in a finite lattice, and 3) far field radiation
properties of the finite lattice. 

\begin{figure} \centering \includegraphics[width=\columnwidth]{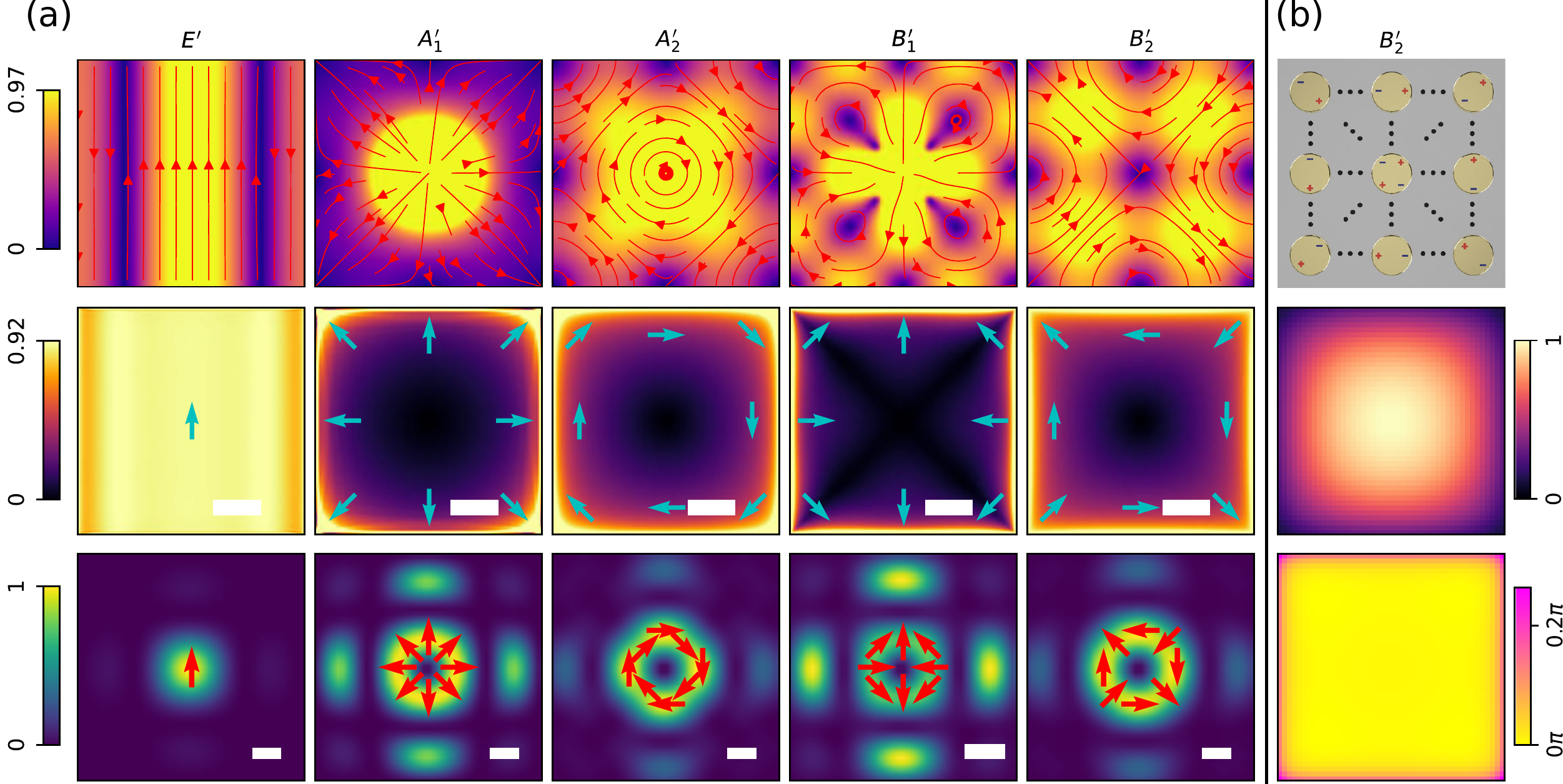}
  \caption{
  (a) Polarizations and electric field patterns corresponding to modes
  from different irreducible representations of the $D_{4h}$ group. Top row:
  Electric field inside a single unit cell (580 nm $\times$ 580 nm) of an
  infinite lattice. The nanoparticle is located in the center of the unit cell.
  Middle row: Magnitude and direction of the induced in-plane dipole moment over
  the entire finite lattice (scale bar \SI{20}{\micro m}). Bottom row: The
  intensity and polarization patterns in the far field produced by the finite
  lattice (scale bar \ang{0.5}). (b) Top: A schematic on the multipole and dipole distributions. 
  Middle and bottom: {Electric quadrupole intensity and phase
  distribution, respectively, in the finite lattice $\B_2'$ mode.} 
   } 
\label{fig:Tmatrixsims} \end{figure}

In the top row of Fig.~\ref{fig:Tmatrixsims} are shown the 5 most prominent
modes for an infinite $D_{4h}$ symmetric lattice derived from group theory: the
degenerate E modes have a well defined polarization in the center of the unit
cell, i.e., at the inversion center of the lattice. For the $\A_1'$, $\A_2'$,
$\B_1'$ and $\B_2'$ modes, however, there exists a polarization singularity at
this point. The topological connection between the polarization singularity in
the momentum space and the diverging Q factor of BICs has been explicitly
pointed out before \cite{SoljacicBIC}. Yet another, perhaps equally intuitive
picture is obtained by studying real space polarization distributions within a
unit cell of the infinite lattice. In particular, the polarization distributions
of BICs are point symmetric, namely the polarization 
$\mathbf{p}(\mathbf{r})=-\mathbf{p}(-\mathbf{r})$, with $\mathbf{r}=0$
being the inversion center (particle center). This directly leads to their
counter-intuitive behaviour: in an infinite lattice, equal magnitude but
opposite polarizations at $\mathbf{r}$ and $-\mathbf{r}$ necessarily result in
destructive interference in the far field normal to the sample (i.e., at the
$\Gamma$ point).  Consequently, this results in an infinite radiative Q factor.
Due to the same reason, the net dipole moment of each unit cell equals zero.

Strictly speaking, genuine BICs are a mathematical construction that are
possible only in infinite systems. It is thus crucial to consider the finite
size effects in any practical implementation. In particular, to what extent can
finite lattices maintain the topological properties of their infinite
counterparts? To study the finite lattice, we employed T-matrix scattering
simulations {in two ways. First, we initially “forcibly” drive the 
nanoparticles inside a \emph{finite} array with multipole excitations 
corresponding to the respective irreps uniformly (as if they were in an 
\emph{infinite} lattice) and observe the fields and induced multipoles 
resulting from the subsequent scattering in the \emph{finite}
array, see Fig.~\ref{fig:Tmatrixsims}(a).
Second, we solve the mode problem in the complex frequency space
\emph{directly for a finite lattice}, see \ref{fig:Tmatrixsims}(b).}

{The middle row of Fig.~\ref{fig:Tmatrixsims}(a) shows
real space dipole moment distributions} along with their polarization
directions for (\SI{100}{\micro m})$^2$ lattices. The E mode intensity is
approximately constant in the center of the lattice, and decreases towards the
edges. This is in agreement with previous observations of the bright mode
intensity distribution \cite{Hoang2017}. In contrast, for the BIC modes the
center of the lattice remains dark due to point symmetric polarization
distribution within the unit cells. At the edges of the structure, however, the
counter-propagating fields that compose the mode are increasingly imbalanced.
This induces a gradual increase of the dipole moment and radiation. Note the
stark contrast to the infinite lattice case, where periodic boundary conditions
render each unit cell to have the same polarization distribution and (zero) net
dipole moment.

Remarkably, the microscopic polarization distribution in each unit cell for the
infinite lattice bears a striking resemblance with the dipole moment
distribution over the entire finite size lattice (compare top and middle row in
Fig.~\ref{fig:Tmatrixsims}(a)). 
As the following arguments apply to each BIC mode,
we inspect only the $\A_1'$ mode in detail. First, in the infinite lattice, the
polarization direction on the right side of the unit cell is the same as the
induced dipole moment on the right edge of the finite lattice.  A closer
inspection indicates this is the case for other directions as well. Second, both
polarization patterns are point symmetric, in the infinite lattice with respect
to the center of the particle and in the finite lattice with respect to the
center of the entire lattice. In the finite lattice case this was further
confirmed by summing 
the electric field at a point (x,y) with the field at (-x,-y) (with (0,0) being
the center of the lattice), see Supplementary Material Fig.~4. This summation
gives zero everywhere in the finite lattice for all BIC modes. Importantly, this
point symmetric polarization distribution in the lattice results in complete
destructive interference in the far field and allows finite lattices to maintain
the infinite radiative Q factor, a feature that has previously been associated
only with infinite lattices.

Third, the polarization exhibits a singularity point in the center of each unit
cell of the infinite lattice. A similar singularity point exists in the center
of the finite lattice. Most importantly, the rotation of the polarization on a
counterclockwise loop around the singularity exhibits a  $2\pi$ counterclockwise
rotation around the center of the particle in the infinite lattice, whereas in
the finite lattice the same applies to the polarization rotation around the
center of the entire lattice. Thus, the microscopic polarization distribution
within each unit cell of the infinite lattice is \textit{imprinted onto the dipole
moment distribution over the entire finite lattice}. 

Next, we inspect the radiation pattern produced by the dipole moment
distribution of the finite lattice. By comparing the middle and bottom row, we
find that the dipole moment distribution in the finite lattice produces
radiation to the far field, whose polarization pattern includes a singularity on
the beam axis and a winding of the polarization that is reminiscent to the
polarization winding at the source plane of the finite lattice. Previously, in
the case of infinite lattices, such far field polarization vortices were
associated with the topological charge, a conserved quantity defined by the
polarization winding number around the singularity
\cite{PhysRevLett.113.257401}. 

Therefore, the following conclusions can be made from 
Fig.~\ref{fig:Tmatrixsims}(a): 
1) In the infinite
lattice, the point symmetric polarization distribution exists in each unit cell.
In the finite lattice, this point symmetry breaks at the single-particle level.
Instead, the point symmetry is preserved with respect to the center of the
entire finite lattice. 2) The resulting dipole moment distribution allows the
BIC to couple coherently to the radiation continuum while still maintaining the
essential topological features of the BICs, namely, destructive interference of
radiation fields normal to the sample (at the $\Gamma$ point), the polarization
singularity and the topological charge associated with the winding number of the
polarization around the singularity.


Remarkably, while BICs have their origin in multipolar resonances
\cite{Grahn2012, Lepetit2014, PhysRevLett.122.153907, Sadrieva2019_2}, it is the
gradual evolution of dipole moments along the finite lattice that governs the
far field intensity and polarization properties of the lasing plasmonic BIC. %
This result also calls for more studies of
finite size effects in photonic crystal structures, wherein BICs have been
observed but finite size effects have only recently been addressed by analytical
(but approximate) envelope function approach \cite{PhysRevB.102.045122}. 

We note that while all BICs produce a donut shape intensity pattern with a
minimum at $\theta_x=\theta_y=0$ ($\Gamma$ point), their polarization properties
are unique. With this insight, we can finally identify the observed BIC mode in
our experiments (Figs.~2-3). First, from the bottom row of
Fig.~\ref{fig:Tmatrixsims}(a) we note that for vertical polarization, $\A_1'$
and $\B_1'$ modes produce two vertically oriented bright lobes, in contradiction
with our experimental data. However, $\A_2'$ and $\B_2'$ produce two
horizontally aligned lobes, similar to the experiments. To distinguish between
these 2 modes, we note that the winding of the polarization around a clockwise
loop along the perimeter of the donut exhibits a 2$\pi$ clockwise rotation for
the $\A_2'$ and a $2\pi$ counterclockwise rotation for the $\B_2'$ mode. Thus, a
clockwise rotating linear polarizer should produce 2 bright lobes, that are
rotating clockwise for the $\A_2'$ and counterclockwise for the $\B_2'$ mode. We
can therefore conclude that the experimentally observed lasing mode in
Fig.~\ref{fig:ffangles} top row is $\B_2'$ carrying a topological charge of -1.

{The observed $\B_2'$ pattern also appears as an actual solution of the finite array mode problem (see Supplementary Material for details). In terms of multipole distributions, the main difference between the actual mode and the results of the uniform multipole driving appears to be the smooth variation in magnitude of the dominant multipole: Fig.~\ref{fig:Tmatrixsims}(b) shows the magnitude and phase distributions of the $(l,m)=(2,-2)$ electric quadrupole dominating in the $\B_2'$ mode. However, the dipole distribution of the mode remains qualitatively the same as the one shown in \ref{fig:Tmatrixsims}(a) for $\B_2'$. Due to computational limitations, the mode simulations were done on a smaller array of size $(\SI{28}{\micro m})^2$, the other parameters being nominally the same as those used in the experiments. The schematic at the top Fig.~\ref{fig:Tmatrixsims}(b) summarizes our findings of the finite size effects. In the center of the lattice, there exists a quadrupolar excitation, similar to the infinite lattice case. This excitation governs the topology of the system and coherently drives the lasing action. Due to finite size effects, however, a gradually increasing dipole moment appears at the edges of the lattice governing the coupling of the subradiant quadrupolar mode to radiation continuum. Remarkably, this coupling maintains the essential topological features of the infinite lattice case.}

\textit{Summary --}
To summarize, we have identified the physical mechanism that allows plasmonic BICs to couple to the radiation continuum. The BIC source plane and far field intensity, coherence, and polarization properties were analyzed in detail. While the BICs have their origin in multipolar resonances, we find that in a finite plasmonic lattice it is the dipole moment distribution that governs the far field properties of the lasing BIC. The dipole moment distribution over the entire finite size lattice bears a striking similarity with the polarization distribution in the unit cell of an infinite lattice. {Consequently, while the point symmetry within each unit cell breaks due to the finiteness of the lattice, the symmetry is maintained with respect to the center of the entire lattice. Importantly, this enables the finite size lattices to maintain the essential topological features of their infinite lattice counterparts.} 




\textbf{Acknowledgements} \par 
Academy of Finland Flagship Programme, Photonics Research and Innovation \textrm{PREIN} 320165, 320166, and 320167; TKH and JH acknowledge Academy of Finland project number 322002, PT and AIV acknowledge project numbers 303351, 327293, 318937
(PROFI). We acknowledge the computational resources provided by the Aalto Science-IT project.\\
\bibliography{references}
\end{document}